\documentclass[12pt]{iopart}
\usepackage{iopams}
\usepackage{mathptmx}

\begin{document}

\title{Duality and conformal twisted boundaries in the Ising model}

\author{Uwe Grimm}

\address{Applied Mathematics Department, 
Faculty of Mathematics and Computing,\\
The Open University, Walton Hall, Milton Keynes MK7 6AA, UK}

\eads{u.g.grimm@open.ac.uk}

\begin{abstract}
There has been recent interest in conformal twisted boundary
conditions and their realisations in solvable lattice models. For the
Ising and Potts quantum chains, these amount to boundary terms that
are related to duality, which is a proper symmetry of the model at
criticality. Thus, at criticality, the duality-twisted Ising model is
translationally invariant, similar to the more familiar cases of
periodic and antiperiodic boundary conditions. The complete
finite-size spectrum of the Ising quantum chain with this peculiar
boundary condition is derived.
\end{abstract}

\address{\it Dedicated to the memory of Sonia Stanciu}

\section{Introduction}

Quantum spin chains are one-dimensional models of interacting quantum
systems. They have been used to model magnetic properties of
materials, in particular when strongly anisotropic behaviour suggests
a one-dimensional modelling. They also arise as limits of
two-dimensional lattice models of classical statistical mechanics, in
an anisotropic limit where the lattice spacing in one space direction
vanishes. In particular, the Ising quantum chain discussed below is
related to the classical Ising model in this way; and the parameter in
the Hamiltonian corresponds to the temperature variable of the
two-dimensional classical model.

At criticality, quantum spin chains as well as two-dimensional lattice
models possess scaling limits that correspond to
$(1\!+\!1)$-dimensional conformal field theories. The critical
exponents describing the non-analytic behaviour of thermodynamic
quantities are given by conformal dimensions of certain conformal
operators. For translationally invariant quantum chains, the scaling
limit corresponds to a conformal field theory on the torus, and the
partition function is a quadratic expression in terms of Virasoro
characters, distinguishing left and right moving excitations.  For
free or fixed boundary conditions, we have a conformal field theory on
the half plane, with a partition function which is linear in Virasoro
characters.

If the spin chain possesses global symmetries, we can define toroidal
boundary con\-di\-tions, i.e., specific twists at the boundary that do
not destroy translational invariance. However, this does not seem to
yield all possible boundary conditions one might expect from the
conformal field theory.  Recently, such ``conformal twisted boundary
conditions'' have attracted growing attention \cite{PZ,CS}, and were
realised in solvable lattice models \cite{CMOP}.  Here, we consider
the simplest possible example of such an exotic boundary condition.
At least in this case, it is once more related to a symmetry of the
model, which turns out to be duality \cite{GS}. Furthermore, the
complete spectrum of the Hamiltonian is obtained exactly \cite{G2},
even for finite chains of length $N$. The scaling limit partition
function is computed, verifying the result expected from a mapping to
the XXZ Heisenberg quantum spin chain which originally led to the
discovery of these duality twisted boundary conditions in the Ising
model \cite{GS}.

Before we come to the particular example of the Ising quantum chain,
we first discuss the construction of toroidal boundary conditions
in more generality.

\section{Quantum Chains and Translational Invariance}

For simplicity, we consider quantum spin chains with nearest neighbour
couplings only. In this case, the Hamiltonian for a system of $N$ spins
has the form
\begin{equation}
H = \sum_{j=1}^{N} H^{}_{j,j+1},\qquad
H^{}_{j,j+1} = \sum_{a,b} \epsilon^{}_{a,b}\sigma_{j}^{a}\sigma_{j+1}^{b}
+ \sum_{a} \delta^{}_{a}\sigma_{j}^{a},
\end{equation}
where $\epsilon^{}_{a,b}$ and $\delta^{}_{a}$ are arbitrary constants
which are independent of $j$.  The Hamiltonian is expressed in terms
of local spin operators $\sigma_{j}^{a}$ acting on a tensor product
space ${\cal V}$,
\begin{equation}
\sigma_{j}^{a} = \mathbf{1}^{\otimes (j-1)}_{V} \otimes \sigma_{}^{a}\otimes
\mathbf{1}^{\otimes (N-j)}_{V}, \qquad
{\cal V} = V^{\otimes N} =
\underbrace{V\otimes V\otimes \ldots \otimes V}_{\mbox{$N$ factors}},
\label{space}
\end{equation}
where $\mathbf{1}_{V}$ denotes the identity operator on the vector
space $V$.  Let us assume periodic boundary conditions for the moment,
i.e., $H^{}_{N,N+1} \equiv H^{}_{N,1}$, so the last spin couples to
the first in the same way as the neighbouring spins along the chain.
We define a unitary translation operator $T$ by its action on the
local spin operators
\begin{equation}
T\sigma_{j}^{a}T^{-1}=\sigma_{j+1}^{a},\; 1\le j\le N-1,\qquad
 T\sigma_{N}^{a}T^{-1}=\sigma_{1}^{a},
\end{equation}
which obviously commutes with the Hamiltonian $H$, so $THT^{-1}=H$.
As $T^{N}=\mathbf{1}$, the identity operator on ${\cal V}$, the
eigenvalues of $T$ are of the form $\exp(2\pi {\rm i}k/N)$, with
$k=0,1,\ldots,N-1$, and define the lattice momenta of the
one-dimensional chain.

Consider now the case where the Hamiltonian has a global symmetry,
i.e., it commutes with an operator $Q=g\otimes g\otimes\ldots\otimes
g=g^{}_{1}g^{}_{2}\ldots g^{}_{N}$, where $g$ belongs to a
representation of some group. We can then define a modified Hamilton
operator $\tilde{H}$ by setting
\begin{equation}
\tilde{H}_{j,j+1} = H_{j,j+1}, \; 1\le j\le N-1,\qquad
\tilde{H}_{N,N+1} \equiv \tilde{H}_{N,1} = g^{}_{1}H_{N,1}g^{-1}_{1},
\end{equation}
so $\tilde{H}$ differs from $H$ only in the coupling term at the
boundary, which is twisted by the local transformation $g$. On first
view, this may appear to be no longer translationally invariant, due
to the different coupling between the first and the last spin.
However, the transformation $Q=g^{}_{1}g^{}_{2}\ldots g^{}_{N}$ is a
symmetry of the model, which means that the nearest neighbour coupling
$H_{j,j+1}$ commutes with the product $g^{}_{j}g^{}_{j+1}$. We thus
can define a modified translation operator
$\tilde{T}=g^{}_{1}T$ which commutes with $\tilde{H}$,
\begin{eqnarray}
\tilde{T}\tilde{H}_{j,j+1}\tilde{T}^{-1}&=&
g^{}_{1}TH_{j,j+1}T^{-1}g^{-1}_{1}=
H_{j+1,j+2}=
\tilde{H}_{j+1,j+2}, \quad
1\le j\le N-1,\nonumber\\
\tilde{T}\tilde{H}_{N-1,N}\tilde{T}^{-1}
&=& g^{}_{1}TH_{N-1,N}T^{-1}g^{-1}_{1}
= g^{}_{1}H_{N,1}g^{-1}_{1}=\tilde{H}_{N,1},\nonumber\\
\tilde{T}\tilde{H}_{N,1}\tilde{T}^{-1}
&=& g^{}_{1}Tg^{}_{1}H_{N,1}g^{-1}_{1}T^{-1}g^{-1}_{1}
= g^{}_{1}g^{}_{2}TH_{N,1}T^{-1}g^{-1}_{2}g^{-1}_{1}\nonumber\\&=&
g^{}_{1}g^{}_{2}H_{1,2}(g^{}_{1}g^{}_{2})^{-1}=H_{1,2}=\tilde{H}_{1,2},
\end{eqnarray}
the corresponding boundary conditions are known as toroidal boundary
conditions.

The Hamiltonian of the Ising quantum chain is given by
\begin{equation}
H_{j,j+1}=-\frac{1}{4}\left(\sigma_{j}^{z}+\sigma_{j+1}^{z}
+2\lambda\sigma_{j}^{x}\sigma_{j+1}^{x}\right),
\qquad
H = -\frac{1}{2} \sum_{j=1}^{N}  \sigma_{j}^{z} +
\lambda\sigma_{j}^{x}\sigma_{j+1}^{x},
\end{equation}
where $\sigma^{x}$ and $\sigma^{z}$ are Pauli matrices, so
$V\cong\mathbb{C }^{2}$. It has global spin reversal symmetry, i.e.,
$H$ commutes with the operator
$Q=\sigma^{z}\otimes\sigma^{z}\otimes\ldots\otimes\sigma^{z}=
\prod_{j=1}^{N}\sigma_{j}^{z}$. Corresponding to this $C_{2}$ symmetry
we have periodic ($H^{\rm P}$ with $g=\mathbf{1}_{V}$) and antiperiodic
($H^{\rm A}$ with $g=g^{-1}=\sigma^{z}_{}$) boundary conditions, the
latter yielding a change in sign of the $\sigma^{x}_{N}\sigma^{x}_{1}$
coupling term, because
$\sigma_{1}^{z}\sigma_{1}^{x}\sigma_{1}^{z}=-\sigma_{1}^{x}$. It turns
out that it is useful to consider the mixed-sector Hamiltonians \cite{BCS}
$H^{+}=H^{\rm P}P_{+}+H^{\rm A}P_{-}$ and $H^{-}=H^{\rm A}P_{+}+H^{\rm
P}P_{-}$ instead, where $P_{\pm}=(\mathbf{1}\pm Q)/2$ are projectors.

\section{Duality Twist in the Ising Quantum Chain}

Duality is a symmetry that relates the ordered and disordered phases
of the classical Ising model. It provides an equality between the
partition functions at two different temperatures, the critical
temperature being mapped onto itself. In the quantum chain language,
duality relates the Hamiltonians $H$ with parameters $\lambda$ and
$1/\lambda$; the critical point corresponds to $\lambda=1$. 

In order to understand the duality transformation, it is advantageous
to rewrite the mixed-sector Hamiltonians $H^{\pm}$ as follows
\begin{equation}
H^{\pm}(\lambda) = 
-\sum_{j=1}^{2N-1} [(e_{2j-1}^{}-{\textstyle\frac{1}{2}})
+\lambda(e_{2j}^{}-{\textstyle\frac{1}{2}})] -
[(e_{2N-1}^{}-{\textstyle\frac{1}{2}})
-\lambda(e_{2N}^{\pm}-{\textstyle\frac{1}{2}})],
\end{equation}
where the Temperley-Lieb operators $e_{j}$ are given by
\begin{equation}
e_{2j-1}=\frac{1}{2}(\mathbf{1}+\sigma^{z}_{j}),\quad
e_{2j}=\frac{1}{2}(\mathbf{1}+\sigma^{x}_{j}\sigma^{x}_{j+1}),\quad
e_{2N}^{\pm}=\frac{1}{2}(\mathbf{1}\pm Q\sigma^{x}_{N}\sigma^{x}_{1}).
\label{eq:ej}
\end{equation}
Defining invertible operators $g_{j}=(1+{\rm i})e_{j}-\mathbf{1}$,
with $g_{j}^{-1}=g_{j}^{\ast}$ and ${\rm i}^2=-1$, the appropriate
duality transformations are $D^{+}=g_{1}g_{2}\ldots g_{2N-1}$ and
$D^{-}=D^{+}\sigma^{x}_{N}$ \cite{GS}. The corresponding duality maps
are $D^{\pm}H^{\pm}(\lambda)=\lambda H^{\pm}(1/\lambda)D^{\pm}$.
Evidently, $D^{\pm}$ act on the operators $e_{j}$ like a translation,
i.e., $D^{\pm}e_{j}=e_{j+1}D^{\pm}$ for $1\le j\le 2N-2$, and at the
boundary $D^{\pm}e_{2N-1}=e_{2N}^{\pm}D^{\pm}$ and
$D^{\pm}e_{2N}^{\pm}=e_{1}D^{\pm}$. Thus the squares of the duality
transformations $D^{\pm}$ commute with the corresponding Hamiltonians
$H^{\pm}$ and are nothing but the appropriate translation operators
$T^{\pm}=(D^{\pm})^{2}$ of the mixed-sector Hamiltonians \cite{GS}.

At criticality, when $\lambda=1$, duality itself becomes a symmetry,
as $D^{\pm} H^{\pm}(1)=H^{\pm}(1)D^{\pm}$. Thus we can define
corresponding twisted boundary conditions. This works in a slightly
different way as for the periodic and antiperiodic boundary conditions
discussed above, as we have to consider an odd number of generators
$e_{j}$. The corresponding mixed-sector Hamiltonians are given by
\cite{GS}
\begin{equation}
\tilde{H}^{\pm}=-\sum_{j=1}^{2N-2}(e_{j}^{}-{\textstyle\frac{1}{2}})
- (e_{2N-1}^{\pm}-{\textstyle\frac{1}{2}}),
\label{eq:ht}
\end{equation}
where the operators $e_{j}$, for $1\le j\le 2N-2$, are defined as in
equation (\ref{eq:ej}) above, and where $e_{2N-1}^{\pm}=(\mathbf{1}\pm
Q\sigma^{y}_{N}\sigma^{x}_{1})/2$. So the duality-twisted Ising
Hamiltonian contains coupling terms of the type
$\pm\sigma^{y}_{N}\sigma^{x}_{1}$ at the boundary, and, in particular,
does {\em not}\/ contain a term $\sigma^{z}_{N}$.

The Hamiltonians $\tilde{H}^{\pm}$ are translationally invariant
\cite{GS}, the corresponding translation operators
$\tilde{T}^{\pm}=(\tilde{D}^{\pm})^{2}$ can be constructed as above as
the squares of the appropriate duality transformations
$\tilde{D}^{+}=g_{1}g_{2}\ldots g_{2N-2}$ and
$\tilde{D}^{-}=\tilde{D}^{+}\sigma^{z}_{N}$, which commute with the
critical Hamiltonians $\tilde{H}^{+}$ and $\tilde{H}^{-}$,
respectively, of equation (\ref{eq:ht}).

\section{Spectrum and Partition Function}

The spectrum of the duality-twisted Ising quantum chain can be
calculated by a modified version \cite{G2} of the standard approach.
Essentially, the Hamiltonians $\tilde{H}^{\pm}$ (\ref{eq:ht}) are
rewritten in terms of fermionic operators by means of a Jordan-Wigner
transformation, and the resulting bilinear expressions in fermionic
operators are subsequently diagonalised by a Bogoliubov-Valatin
transformation, see \cite{G2} for details. The diagonal form of the
Hamiltonian is
\begin{equation}
\tilde{H}^{\pm} = \sum_{k=0}^{N-1}\Lambda_{k}
\eta_{k}^{\dagger}\eta_{k}^{}+ E_{0}\mathbf{1}
\end{equation}
where $\eta_{k}^{\dagger}$ and $\eta_{k}$ are fermionic creation and
annihilation operators, respectively. The energies of the elementary
fermionic excitations are given by
\begin{equation}
\Lambda_{k} = 2|\sin(\case{p_{k}}{2})|,\qquad
p_{k} = \frac{4k\pi}{2N-1},\qquad k=0,1,2,\ldots,N-1.
\end{equation}
The ground-state energy $E_{0}$ is 
\begin{equation}
-E_{0}^{}=
\sum_{k=0}^{N-1}\sin(\case{k\pi}{\tilde{N}})=
\frac{1+\cos(\case{\pi}{2\tilde{N}})}{2\sin(\case{\pi}{2\tilde{N}})}
=\frac{2\tilde{N}}{\pi}-\frac{\pi}{24(\tilde{N})}+\Or[(\tilde{N})^{-3}]
\end{equation}
which shows the expected finite-size corrections of a translational
invariant critical quantum chain with an effective number of sites of
$\tilde{N}=N-1/2$, reminiscent of the fact that it is related to the
XXZ Heisenberg quantum chain with an odd number $2\tilde{N}=2N-1$ of
sites \cite{GS}.  With the appropriate finite-size scaling, the
linearised low-energy spectrum in the infinite system is
\begin{equation}
\frac{\tilde{N}}{2\pi}\,
(\tilde{H}^{\pm}-\frac{\tilde{N}}{N}E_{0}^{\rm P}\mathbf{1})
\;\stackrel{\scriptscriptstyle N\rightarrow\infty}{\longrightarrow}\;
\sum_{r=0}^{\infty}[ra_{r}^{\dagger}a_{r}^{} +
(r+\case{1}{2})b_{r}^{\dagger}b_{r}^{}]+\frac{1}{16}
\end{equation}
where the fermionic operators $a_{k}$ and $b_{k}$ follow from the
$\eta_{k}$ by suitable renumbering, and where $E_{0}^{\rm
P}=1/\sin(\frac{\pi}{2N})$ denotes the ground-state energy of the
$N$-site Ising quantum chain with periodic boundary conditions. The
conformal partition functions are given by the combinations
$(\chi_{0}+\chi_{1/2})\bar{\chi}_{1/16}$ and
$\chi_{1/16}(\bar{\chi}_{0}+\bar{\chi}_{1/2})$, respectively, of
characters $\chi_{\Delta}$ of irreducible representations with highest
weight $\Delta$ of the $c=1/2$ Virasoro algebra, corresponding to
operators with conformal spin $1/16$ and $7/16$.

\section{Concluding Remarks}

For the Ising and Potts quantum chains \cite{GS}, ``exotic'' conformal
twisted boundary conditions can be realised by means of twists related
to duality, which is a symmetry of the model at criticality. It would
be interesting to know whether this is a more general feature. If this
is the case, this observation might help to identify non-trivial
symmetries in quantum chains or two-dimensional solvable lattice
models of statistical mechanics.

\Bibliography{9}

{\normalsize

\bibitem{PZ}
Petkova V B and Zuber J-B 2001
Generalised twisted partition functions
{\it Phys.\ Lett.}\ B {\bf 504} 157--64 

\bibitem{CS} 
Coquereaux R and Schieber G 2001 
Twisted partition functions for ADE boundary conformal field theories
and Ocneanu algebras of quantum symmetries
{\it J.\ Geom.\ Phys.}\ {\bf 42} (2002) 216--258

\bibitem{CMOP}
Chui C H O, Mercat C, Orrick W P and Pearce P A 2001
Integrable lattice realizations of conformal twisted boundary conditions
{\it Phys.\ Lett.}\ B {\bf 517} 429--35 

\bibitem{GS}
Grimm U and Sch\"{u}tz G 1993
The spin-1/2 XXZ Heisenberg chain, the quantum algebra U$_q$[sl(2)], 
and duality transformations for minimal models
{\it J.\ Stat.\ Phys.}\  {\bf 71} 921--64

\bibitem{G2}
Grimm U 2002
Spectrum of a duality-twisted Ising quantum chain
{\it J. Phys. A: Math. Gen.} {\bf 35} (2002) L25--30

\bibitem{G1}
Grimm U 1990
The quantum Ising chain with a generalized defect
{\it Nucl.\ Phys.}\ B {\bf 340} 633--58

\bibitem{BCS}
Baake M, Chaselon P and Schlottmann M 1989
The Ising quantum chain with defects. II.~The so($2n$) Kac-Moody spectra
{\it Nucl.\ Phys.}\ B {\bf 314} 625--45
}
\endbib

\end{document}